\documentstyle[prl,aps,multicol,epsfig,amssymb]{revtex}

\begin{document}

\draft

\title{Velocity correlations in dense granular gases}
\author{Daniel L. Blair and A. Kudrolli} 
\address{Department of Physics, Clark University, Worcester, MA 01610, USA}
\date{\today}
\maketitle
	
\begin{abstract}

We report the statistical properties of spherical steel
particles rolling on an inclined surface being driven by an
oscillating wall. Strong dissipation occurs due to collisions
between the particles and rolling and can be tuned by changing the
number density. The velocities of the particles are observed
to be correlated over large distances comparable to the system
size. The distribution of velocities  deviates strongly from a
Gaussian. The degree of the deviation, as measured by the kurtosis
of the distribution, is observed to be as much as four times the
value corresponding to a Gaussian, signaling a significant
breakdown of the assumption of negligible velocity correlations
in a granular system.
\end{abstract}
\begin{multicols}{2}
From the study of landslides to the understanding of the formation
of galaxies, systems of macroscopic particles are of fundamental
interest for physicists and engineers~\cite{jnb,astro}. To treat
a granular system statistically, temperature must be replaced by a
more suitable quantity. Granular temperature has gained
considerable attention over the past two decades as a useful
quantity for describing the properties of rapid granular
flow~\cite{ogawa78,haff83,jenkins83}. If the collisions between
particles are elastic, then we would expect on very general grounds
that the velocity distribution is Gaussian and the velocity
components are uncorrelated. However, dissipation due to
inelasticity and friction can introduce correlations that makes
these properties of the velocity distribution
suspect~\cite{collapse}. For example, a simulation of a system of
inelastic particles yields long-range velocity
correlations~\cite{bizon}.

The first experiments with steel particles excited by a vertically
vibrated container indicated that the velocity distribution is
Gaussian\cite{warr95}. Using advances in high speed digital imaging,
later experiments have shown deviations in the tails of the
distribution from a Gaussian~\cite{losert99,olafsen99,rouyer00}. It
also has  been claimed that the distribution of the velocity
component, $v_x$, in the direction perpendicular to the direction of
the vibration, may be described by the functional form $R(v_x) =
\exp[-(|v_x|/\sigma_x)^{3/2}]$, where $\sigma_x =
{\left<v_{x}^2\right>}^{1/2}$~\cite{rouyer00,pug_pre}. However,
these experiments were conducted with particles that are
relatively elastic, and the velocity
correlations were either not measured or claimed to be negligible.
Particles that collide while rolling lose more energy in
comparison to particles that collide in free space, due to
frustration at impact. In earlier work~\cite{kudrolli00}, this
effect lead to strong deviations even in a dilute system where the
mean free path was of the order of the dimension of the container. 
However, dissipation at the boundary was significant and the
distribution did not represent the bulk properties of the
particles. These observations lead us to ask if there are
correlations in dense systems where the mean free path is much 
smaller than the system size, and if these correlations affect the
distribution of the velocities.

In this Letter, we address these questions by measuring the
velocities of particles rolling on an
inclined plane. This geometry slows the dynamics and allows us to
vary the dissipation over a wide range. Energy is supplied to the
system by means of an oscillating wall. We report the density and
velocity distributions as a function of the number of particles
$N$ in a regime where the distributions are insensitive to the
frequency or phase of the drive. The dissipation increases with
$N$ because the time between collisions as well as  the mean free
path decreases. Because the density of the particles becomes
higher near the driving wall as $N$ is increased, our analysis of
the velocities is done in a narrow region where the density and
the velocity fluctuations are relatively uniform.

We observe that significant correlations in the velocities of the
particles that increase in strength and range as
$N$ increases are found. The distribution,
$P(v_x)$, of the velocity components of the particles
perpendicular to the direction of the oscillation deviates
strongly from a Gaussian. At low $N$, corresponding to lower
dissipation, $P(v_x)$ may be fitted to $R(v_x)$ as in
Refs.~\cite{rouyer00,pug_pre}, but
$P(v_x)$ deviates strongly from this form as $N$ is increased. 
The kurtosis of $P(v_x)$ is observed to be as much as {\it four
times} the value corresponding to a Gaussian distribution,
signaling a significant breakdown of the assumptions of negligible
velocity correlations in our system.

The experimental system consists of stainless steel spheres with
diameter $d=0.3175\,\rm cm$ rolling on a $29.5\,{\rm cm \times
27.0\, cm}$ glass plane that can be tilted from $0 = \theta \leq
8^\circ$; a schematic of the experimental setup is given in
Ref.~\cite{kudrolli00}. The coefficient of rolling friction is
observed to be less that $1
\times 10^{-3}$, and therefore energy loss during rolling is
negligible compared to the loss that occurs due to a
collision~\cite{kondic99,paint00}.  The number of particles is
varied from $N = 100 -1000$.   To track the positions of the
particles we utilize a high speed digital camera (Kodak
Motioncorder) at a frame rate of 250 per second. A
centroid technique is used to follow the particles at sub-pixel
accuracy, allowing us to resolve positions to within 0.13\,mm and
velocities to within
$5.2 \times 10^{-4}\,{\rm cm\, s}^{-1}$. The statistical
properties are obtained by averaging over time using at least 8500
frames.

\begin{figure}[tb]
\centerline{\epsfig{file=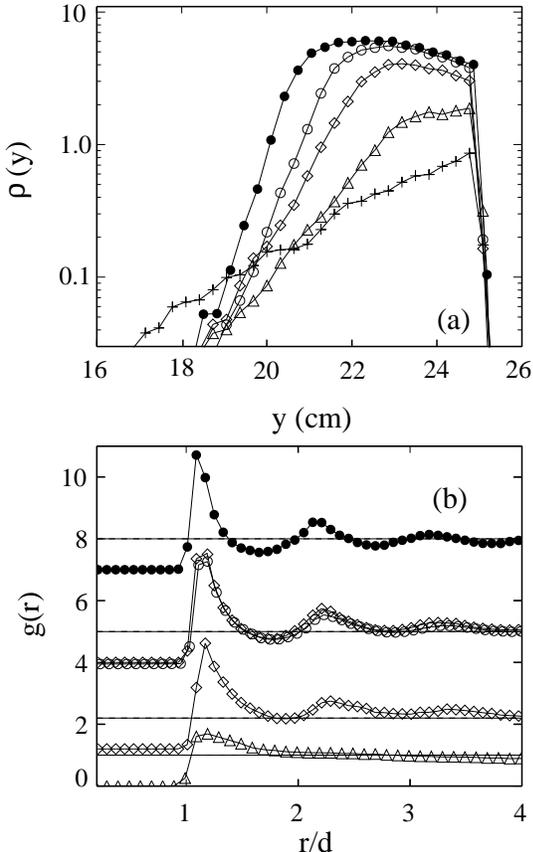,width=7.0cm}}
\caption{(a) The average number density, $\rho(y)$, in the
$y$-direction for various values of $N$; 1000 ($\bullet$), 750
($\circ$), 500 ($\diamond$), 200 ($\vartriangle$), 100 ($+$).  The
driving wall is at $y = 26\,\rm cm$.  (b) The radial distribution
function $g(r)$ for the same values of $N$ at $f = 6$\,Hz ($d$ is
the particle diameter). The plots of $g(r)$ have been shifted for
clarity, and the dashed lines show
$g(r)=1$. To show that the details of the driving do not
influence the correlation of the particles, 
$g(r)$ for
$N=750$ at
$f=6$\,Hz ($\circ$) and $f=20$\,Hz ($\diamond$) are shown.}
\label{system}
\end{figure}

Energy is added to the system by means of the bottom wall which is
attached to a solenoid powered by a waveform generator. The driving
signal is not sinusoidal, but a periodic pulse with an amplitude
${\cal A} \approx 2.5\,d$ and a velocity  $v_{\rm wall} \approx
20\,{\rm cm\ s}^{-1}$. When the system is tilted past $\theta =$
2.0$^\circ$ and the driving frequency  $f \gtrsim 4.0\,$Hz, the
positions of the particles are observed to be independent of the phase
of the drive. Below 4.0\,Hz and for $N > 200$, we observe particles
forming subharmonic patterns similar to those observed in vertical
vibrated quasi-two-dimensional geometries~\cite{douady89}. The nature
of the velocity fields and density distribution in this latter regime
will be discussed elsewhere~\cite{blair01}. All of our results
reported here are at a fixed  $\theta = 3.0^\circ$, ${\cal A} =
2.5\,d$, and $f = 6.0\,\rm Hz$. For these parameters phase dependence
is absent in the density and velocity distributions.

\begin{figure}[tb]
\centerline{\epsfig{file=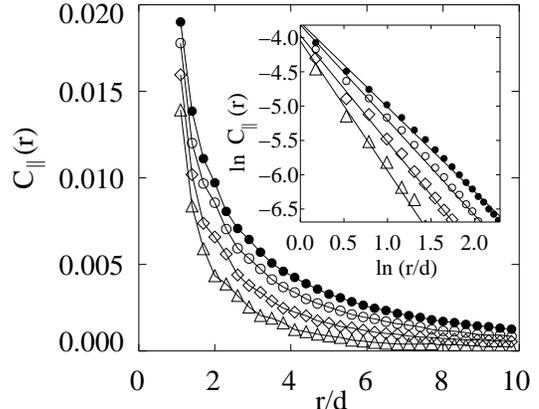,width=7 cm}}
\caption{The spatial correlation function of the velocities,
$C_{||}(r)$, as a function of separation $r/d$, for $N= 1000$
($\bullet$), 750 ($\circ$), 500 ($\diamond$), and 200
($\vartriangle$). {\em Inset: } To show the nonexponential decay of
$C_{||}(r)$, we plot $\ln C_{||}(r)$ versus $\ln\left(r/d\right)$. The
solid lines are least squares fits to the data. The slope $\alpha=
-1.23$ ($\bullet$), $-1.38$ ($\circ$), $-1.55 $ ($\diamond$),
$-1.90$($\vartriangle$). The statistical error bars are smaller then
the symbols.} \label{vel_corr}
\end{figure}

The $x$ and $y$ axes are chosen to be along the driving wall and
perpendicular to the driving wall respectively. The origin corresponds
to the point where the top wall intersects the side wall. The
distribution of the particles in the $x$-direction is uniform. The
$y$-dependence of the number density, $\rho(y)$, is obtained by
averaging the number of particles in bins of width $d$ and dividing by
the area of each bin (see Fig.~\ref{system}(a)).  The density
$\rho(y)$ is not constant because of gravity and the action of the
driving wall, although for systems with $N > 100$, $\rho(y)$ shows a
plateau where there are very small density gradients, similar to
earlier observations~\cite{luding94,kudrolli97}. Most of the particles
remain suspended above the driving wall and receive random kicks from
the particles that are between the bulk and the driving wall.

To ensure that the deviations in density over the compacted region
do not influence the distribution of velocities, we varied the
region over which the analysis of the velocities was performed. We
found that the distribution of velocities is independent of the
distance $y$ in the compacted region. Therefore, we measured the
velocity distributions in a region of width $5d$ beginning
$8d$ away from the driving wall.

To further characterize the compacted region in our experiment, we
measure the  radial distribution function $g(r)$ as a function of the
distance $r$ between the centers of the particles~\cite{bizon}. Our
results for various $N$ are shown in Fig.~\ref{system}(b). The
plots demonstrate a transition from a gas-like state for  $N \leq
200$, to a more liquid-like state for $N \gtrsim 500$.

We now discuss the detailed properties of the velocities of the
particles. The spatial correlation function of the velocities is
given by
\begin{equation}\vspace{-.2in}
C_{||}(r) = \sum_{i\ne j} \frac{\vec{v_i}\cdot \vec{v_j}}{|\vec
v_i||\vec v_j|},
\vspace{-.2in}
\label{corr}
\end{equation}
where $i$ and $j$ label particles separated by a distance $r$. With
this definition, two particles with parallel (antiparallel)
velocities give a correlation of $+1$ ($-1$). We find that
long-range velocity correlations are present for all $N$ as shown
in Fig.~\ref{vel_corr}.  We have  plotted $\ln C_{||}(r)$ versus
$\ln\,(r/d)$ in the inset of Fig.~\ref{vel_corr}. The
$r$-dependence is nonexponential, and least squares fits to $\ln
C_{||}(r)$ suggest a possible power law decay. However, our range
of $r$ is limited by the finite size of our system. The values of
the slope $\alpha$ are  given in the caption of
Fig.~\ref{vel_corr}.

\begin{figure}[tb]
\centerline{\epsfig{file=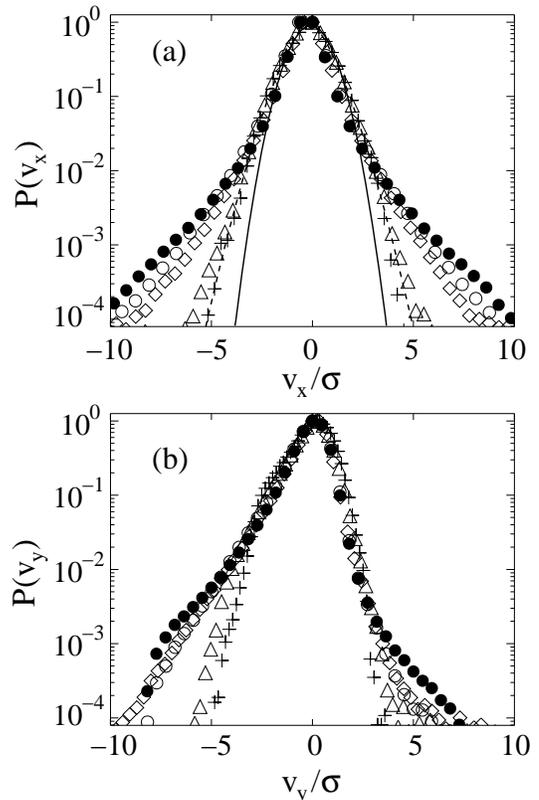,width=7cm}}
\caption{The distribution of velocities for $N = 1000$ ($\+ \bullet$),
$750$ ($\+ \circ$), $500$ ($\+ \diamond$), $200$ ($\+ \vartriangle$),
and $100$ ($\+ +$).  (a) The velocity distribution $P(v_x)$
for the region described in the text.  Each
distribution is normalized to be unity at its maximum.  The
deviation from a Gaussian (solid line) are
significant for all $N$ and show a systematic increase as a
function of $N$.  In the most dilute case ($+$) $P(v_x)$
can be described by $R(v_x) =
\exp[-(|v_x|/\sigma_x)^{3/2}]$ shown by the dashed line. (b) The
distributions $P(v_y)$ are observed to be asymmetric due to
the driving wall and dissipation. The asymmetry is observed to
increase as $N$ is increased. However, this asymmetry does not cause
the deviations from a Gaussian in $P(v_x)$ (see text).}
\label{vel_dist} 
\end{figure}

We now discuss how these correlations affect the distribution of
velocities in the $x$ and $y$ directions (see
Figure~\ref{vel_dist}).  The velocity
components $v_x$ and $v_y$ have been scaled by $\sigma_x =
\left<v_x^2\right>^{1/2}$ and
$\sigma_y =
\left<v_y^2\right>^{1/2}$. The maximum of the distribution $P(v_x)$ is
scaled to be unity for clarity and is given in Fig.~\ref{vel_dist}(a)
for various $N$. A Gaussian fit is shown for comparison. For low
densities, we find that the form of $P(v_x)$ is similar to that
observed in previous experiments with low
dissipation~\cite{losert99,rouyer00}. For $N = 100$
($+$), the form of $P(v_x)$ can be fitted by $R(v_x) =
\exp[-(|v_x|/\sigma_x)^{3/2}]$, shown by the dashed line. However, as
$N$ is increased, the deviations are observed to become stronger and
cannot be described by this power law form.

The asymmetry of $P(v_y)$ seen in Fig.~\ref{vel_dist}(b) occurs
because the particles moving toward the piston have lost energy
because of collisions. The distributions become more asymmetric as
$N$ is increased because of the greater number of collisions
between particles. To characterize $P(v_y)$, we
calculate its skewness (third moment), and find that it ranges
from $1.62-4.45$ for
$N=100-1000$ respectively. The effects of this asymmetry on
$P(v_x)$ can be determined by discriminating particles with
$v_y<0$.  After performing this ``filtering,'' we find that
$P(v_x)$ is insensitive to the sign of
$v_y$. Thus the asymmetry of $P(v_y)$ is not the cause of the
deviations in $P(v_x)$ from a Gaussian.

\begin{figure}[tb]
\centerline{\epsfig{file=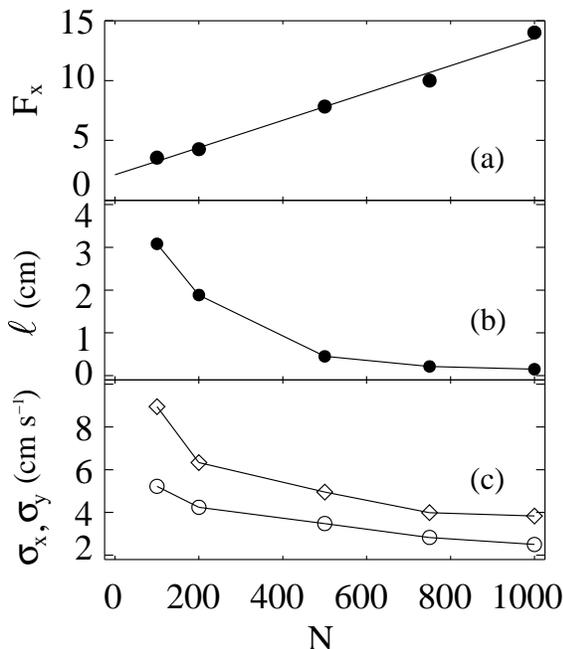,width=7.5cm}}
\caption{(a) The kurtosis $F_x$ of $P(v_x)$, as a function of $N$
($F=3$ for a Gaussian.)  The  solid line is a least squares fit to the
data. (b) The mean free path $\ell$ as function of the number of
particles $N$.  (c) The standard deviations $\sigma_x$ ($\circ$)
and
$\sigma_y$ ($\vartriangle$) (measure of granular temperature)  as a
function of $N$, taken from the distributions in Fig.~\ref{vel_dist}.
The difference of $\sigma_x$ and $\sigma_y$ reiterates that the
granular temperature is a nonscalar quantity.  The line is
a guide for the eye.}
\label{flat}
\end{figure}

By using the kurtosis or flatness of the distribution $P(v_x)$ given
by $F_x = \left<v_{x}^4\right>/\left<v_{x}^2\right>^2$,  we can
quantify the deviation of $P(v_x)$ from a Gaussian.  If the
distributions were Gaussian, $F_x = 3.0$. If the distributions
were given by $R(v_x)$, $F_x = 3.762$.  Figure~\ref{flat}(c)
shows $F_x$ as a function of $N$. The increase in $F_x$ shown in
Fig.~\ref{flat}(a) implies that $P(v_x)$  deviates more strongly from
a Gaussian as dissipation is increased. The  linear dependence of
$F_x$ on $N$ may be due to the finite range of $N$.

These results show that there are strong correlations, both in the
positions and in the velocities, as a consequence of the high
dissipation that occurs due to collisions. If steel particles collide
head-on in free space, they lose energy as determined by the
coefficient of normal restitution given by $\eta \approx 0.93$.
However, when rolling particles collide, the translational velocity
changes at the instant of collision, but the angular velocity does
not. Therefore, the rolling condition is no longer satisfied and the
particles slide for a short duration. Because the sliding friction is
about 100 times larger than the rolling friction for steel on glass,
the rolling condition is established within a few particle
diameters~\cite{kudrolli00,kondic99,paint00}.  Hence, particles
lose more energy than they would by considering only the normal
coefficient of restitution. Measurements for our system show that
the effective coefficient of restitution $\eta_{\rm eff} \approx
0.5$~\cite{kudrolli00}. Because the effect of inelasticity is to
reduce relative motion~\cite{inelastic}, this smaller value implies
that the particles tend to cluster and stream when driven, leading
to the observed spatial and velocity correlations.

To estimate the increase in dissipation with $N$, we calculate the
mean free path $\ell$ from the relation given in
Ref.~\cite{grossman97}. The results for $\ell$ as a function of $N$
are shown in Fig.~\ref{flat}(b).  We find that $\ell$ decreases
much faster than the estimate based on an uniform spatial
distribution of particles. By dividing $\ell$ by the root mean
square velocity, we obtain the mean collision time between
particles.  This time is observed to decrease from 0.83\,s to
0.059\,s as $N$ is increased. The increase in the range of the
velocity correlations and the deviations from a Gaussian in
$P(v_x)$ is due to the higher dissipation in the system for larger
$N$.  As $N$ is increased, the collision rate increases, which
gives rise to higher dissipation. As the latter is increased, the
overall particle velocities decrease. This decrease in turn leads
to more compaction and smaller mean free paths.

To ensure that the means of energy input is not effecting the
system, we have varied the driving parameters.  We find that the
statistical properties do not depend strongly on how the system is
driven above a threshold  $f \gtrsim 4\,$Hz. Increasing $f$ and
the amplitude
$\cal A$ of the oscillating wall  by a factor of five has no
effect on the spatial and velocity distributions for fixed $N$.
For example, Fig.~\ref{system}(b) shows that for $N=750$, $g(r)$
remains unchanged if $f$ is increased by a factor of 3.3  and
$\cal A$ is increased by a factor of 3.0.

In Fig.~\ref{flat}(c) we show $\sigma_x$  and $\sigma_y$ as a function
of $N$ to confirm the anisotropy of the granular temperature for our
system. The plot also shows that the average energy per particle
decreases as $N$ is increased. This decrease occurs due to the higher
collision rate between the particles that removes energy from the
system through dissipation.

In summary, strong velocity correlations in dense granular gases
have been experimentally reported for the first time.  These
correlations are observed to cause the velocity distributions to
deviate significantly from a Gaussian. Our experiments also
suggest that the velocity distributions  depend on the degree of
dissipation in the system and the form
$R(v_x)=\exp[-(|v_x|/\sigma_x)^{3/2}]$ of the velocity distribution
is not universal.

We thank H.~Gould and J. Tobochnik for fruitful conversations,
E.~Weeks and J.~Crocker for their routines to detect particle
positions, and J.~Norton for technical assistance. This project was
supported by the donors of the Petroleum Research Fund, and the
National Science Foundation under Grant \# DMR-9983659. A.~K.\ was
also supported by a fellowship from the Alfred P.~Sloan Foundation.

\end{multicols}
\end{document}